\documentclass[12pt]{iopart}

\expandafter\let\csname equation*\endcsname\relax
\expandafter\let\csname endequation*\endcsname\relax
\usepackage{amsmath}
\usepackage{amsthm}
\usepackage{iopams}
\usepackage{bbm}
\usepackage{braket}
\usepackage{graphicx}
\usepackage{epstopdf}
\usepackage{enumitem}

\newenvironment{myenumerate}{%
  \edef\backupindent{\the\parindent}%
  \enumerate%
  \setlength{\parindent}{\backupindent}%
}{\endenumerate}

\usepackage{color,soul}
\sethlcolor{yellow}

\theoremstyle{plain}
\newtheorem{theorem}{Theorem}
\theoremstyle{plain}
\newtheorem{lemma}{Lemma}

\newcommand{\wt}{\widetilde}

\def\>{\rangle}
\def\<{\langle}
\def\E{ {\cal E} }
\def\P{ {\cal P} }
\def\EA{{\cal E}_A}
\def\EB{{\cal E}_B}
\def\E{\cal E}
\def\EBA{{\cal E}_{B|A}}
\def\EAB{{\cal E}_{A|B}}
\def\EBC{{\cal E}_{B|C}}
\def\ECB{{\cal E}_{C|B}}

\def\P{ {\cal P} }

\def\I{ \mathbbm{1} }

\DeclareMathOperator{\diag}{diag}

\def\diag{ \mbox{diag} }

\setenumerate[0]{label=(\alph*)}

\begin{document}

\title[Quantum steering ellipsoids, extremal physical states and monogamy]{Quantum steering ellipsoids, extremal physical states and monogamy}

\author{Antony Milne$^1$, Sania Jevtic$^2$, David Jennings$^1$, Howard Wiseman$^3$ and Terry Rudolph$^1$}

\address{$^1$ Controlled Quantum Dynamics Theory, Department of Physics, Imperial College London, London SW7 2AZ, UK}
\address{$^2$ Mathematical Sciences, Brunel University, Uxbridge UB8 3PH, UK}
\address{$^3$ Centre for Quantum Computation and Communication Technology (Australian Research Council), Centre for Quantum Dynamics, Griffith University, Brisbane, Queensland 4111, Australia}

\ead{\mailto{antony.milne@gmail.com}}

\begin{abstract}Any two-qubit state can be faithfully represented by a steering ellipsoid inside the Bloch sphere, but not every ellipsoid inside the Bloch sphere corresponds to a two-qubit state. We give necessary and sufficient conditions for when the geometric data describe a physical state and investigate maximal volume ellipsoids lying on the physical-unphysical boundary. We derive monogamy relations for steering that are strictly stronger than the Coffman-Kundu-Wootters (CKW) inequality for monogamy of concurrence. The CKW result is thus found to follow from the simple perspective of steering ellipsoid geometry. Remarkably, we can also use steering ellipsoids to derive non-trivial results in classical Euclidean geometry, extending Euler's inequality for the circumradius and inradius of a triangle.\end{abstract}


\maketitle

\section{Introduction}

The Bloch vector representation of a single qubit is an invaluable visualisation tool for the complete state of a two-level quantum system. Properties of the system such as mixedness, coherence and even dynamics are readily encoded into geometric properties of the Bloch vector. The extraordinary effort expended in the last 20 years on better understanding quantum correlations has led to several proposals for an analogous geometric picture of the state of two qubits~\cite{Avron2007,Bengtsson2006,Horodecki1996}. One such means is provided by the quantum steering ellipsoid~\cite{Verstraete2002,Shi2011,Jevtic2013,Altepeter09}, which is the set of all Bloch vectors to which one party's qubit could be `steered' (remotely collapsed) if another party were able to perform all possible measurements on the other qubit. 

It was shown recently~\cite{Jevtic2013} that the steering ellipsoid formalism provides a faithful representation of all two-qubit states and that many much-studied properties, such as entanglement and discord, could be obtained directly from the ellipsoid. Moreover steering ellipsoids revealed entirely new features of two-qubit systems, namely the notions of complete and incomplete steering, and a purely geometric condition for entanglement in terms of nested convex solids within the Bloch sphere. 

However, one may well wonder if there is much more to be said about two-qubit states and whether the intuitions obtained from yet another representation could be useful beyond the simplest bipartite case. We emphatically answer this in the affirmative. Consider a scenario with three parties, Alice, Bob and Charlie, each possessing a qubit. Bob performs measurements on his system to steer Alice and Charlie. We show that the volumes $V_{A|B}$ and $V_{C|B}$ of the two resulting steering ellipsoids obey a tight inequality that we call the \emph{monogamy of steering} (Theorem \ref{volume_bounds}):\begin{eqnarray}\sqrt{V_{A|B}}+\sqrt{V_{C|B}}\le \sqrt{\frac{4\pi}{3}}.\end{eqnarray}

We also prove an upper bound for the concurrence of a state in terms of the volume of its steering ellipsoid (Theorem \ref{concurrence_vol}). Using this we show that the well-known CKW inequality for the monogamy of concurrence~\cite{Coffman2000} can be derived from the monogamy of steering. The monogamy of steering is therefore strictly stronger than the CKW result, as well as being more geometrically intuitive.

The picture that emerges, which was hinted at in Ref.~\cite{Jevtic2013} by the nested tetrahedron condition for separability, is that the volume of a steering ellipsoid is a fundamental property capturing much of the non-trivial quantum correlations. But how large can a steering ellipsoid be? Clearly the steering ellipsoid cannot puncture the Bloch sphere. However, not all ellipsoids contained in the Bloch sphere correspond to physical states. We begin our analysis by giving necessary and sufficient conditions for a steering ellipsoid to represent a valid quantum state (Theorem \ref{conditions}). The conditions relate the ellipsoid's centre, semiaxes and orientation in a highly non-trivial manner.

We subsequently clarify these geometric constraints on physical states by considering the limits they impose on steering ellipsoid volume for a fixed ellipsoid centre. This gives rise to a family of extremal volume states (Figure \ref{fig:maximal_volume_ellipsoids}) which, in Theorem \ref{indicator}, allows us to place bounds on how large an ellipsoid may be before it becomes first entangled and then unphysical. The maximal volume states that we give in equation \eqref{eq:max_state} are found to be very special. In addition to being Choi-isomorphic to the amplitude-damping channel, these states maximise concurrence over the set of all states that have steering ellipsoids with a given centre (Theorem \ref{max_vol_conc}). This endows steering ellipsoid volume with a clear operational meaning.

A curious aside of the steering ellipsoid formalism is its connection with classical Euclidean geometry. By investigating the geometry of separable steering ellipsoids, in Section \ref{section_geometry} we arrive at a novel derivation of a famous inequality of Euler's in two and three dimensions. On a plane, it relates a triangle's circumradius and inradius; in three dimensions, the result extends to tetrahedra and spheres. Furthermore, we give a generalisation of Euler's result to ellipsoids, a full discussion of which appears in Ref.~\cite{GeometryPaper}.

The term `steering' was originally used by Schr\"odinger~\cite{Schrodinger1935} in the context of his study into the complete set of states/ensembles that a remote system could be collapsed to, given some (pure) initial entangled state. The steering ellipsoid we study is the natural extension of that work to mixed states (of qubits). Schr\"odinger was motivated to perform such a characterisation by the EPR paper~\cite{EPR}. The question of whether the ensembles one steers to are consistent with a local quantum model has been recently formalised~\cite{Wiseman2007} into a criterion for `EPR steerability' that provides a distinct notion of nonlocality to that of entanglement: the EPR-steerable states are a strict subset of the entangled states. We note that the existence of a steering ellipsoid with nonzero volume is necessary, but not sufficient, for a demonstration of EPR-steering. It is an
 open question whether the quantum steering ellipsoid can provide a geometric intuition for EPR-steerable states as it can for separable, entangled and discordant states, although progress has recently been made~\cite{UnpublishedJevtic}.

\begin{figure}
\begin{indented}
\item[]\includegraphics[width=0.4\textwidth]{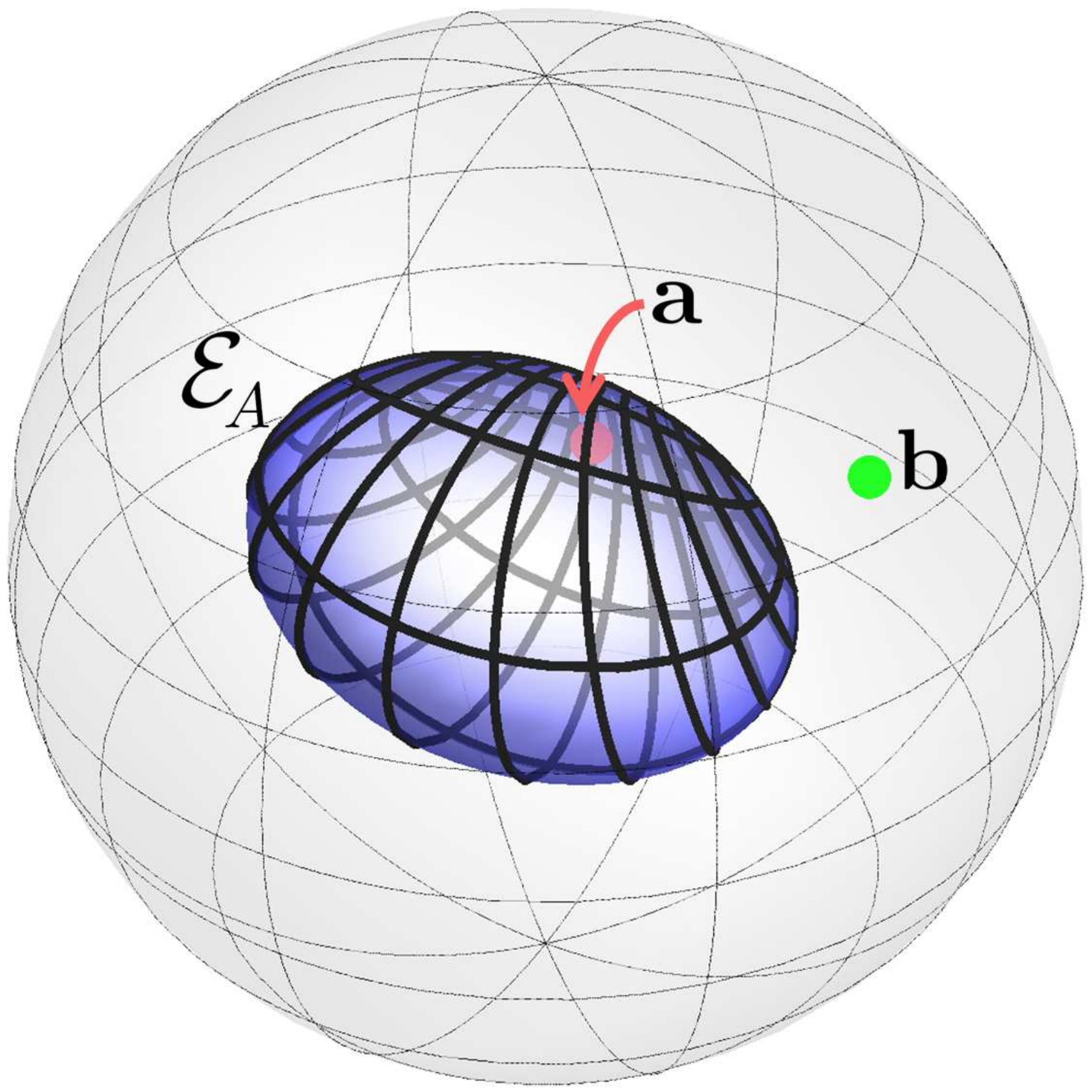}
\end{indented}
\caption{An example of the geometric data: Alice's steering ellipsoid $\EA$ and the two Bloch vectors $\bi a$ and $\bi b$. Together with a specification of Bob's local basis, this is a faithful representation of two-qubit states.}
\label{fig:generic_ellipsoid}
\end{figure}

\section{The canonical, aligned state}\label{section_canonical}

A Hermitian operator with unit trace acting on the Hilbert space $\mathbb{C}^2\otimes\mathbb{C}^2$ may be expanded in the Pauli basis $\{\mathbbm{1}, \bsigma\}^{\otimes 2}$ as
\begin{eqnarray}\label{eq:general_state}
\rho=\frac{1}{4}(\mathbbm{1}\otimes\mathbbm{1}+\bi{ a}\cdot\bsigma\otimes\mathbbm{1}+\mathbbm{1}\otimes\bi{b}\cdot\bsigma+\sum_{i,j=1}^3T_{ij}\,\sigma_i\otimes\sigma_j).
\end{eqnarray}
For a two-qubit state, $\rho$ is positive semidefinite, $\rho\geq 0$. The local Bloch vectors are given by $\bi{a}=\tr(\rho\,\bsigma \otimes \mathbbm{1})$ and $\bi{b}=\tr(\rho\,\mathbbm{1}\otimes \bsigma)$, whilst bipartite correlations are contained in the matrix $T_{ij}=\tr(\rho \,\sigma_i \otimes \sigma_j)$~\cite{Horodecki1996}. Requiring that $\rho \geq 0$ places non-trivial constraints on $\bi{a}$, $\bi{b}$ and $T$.

Alice's steering ellipsoid $\EA$ is described by its centre 
\begin{eqnarray}
\bi{c}_A =\gamma_b^{2}(\bi a - T \bi b),
\label{eq:centreA}
\end{eqnarray}
where the Lorentz factor $\gamma_b=1/\sqrt{1-b^2}$, and a real, symmetric $3\times 3$ matrix 
\begin{eqnarray}
Q_A =\gamma_b^{2}\left(T-\bi a \bi b^\mathrm{T} \right)\left( \I + \gamma_b^2 \bi b \bi b^\mathrm{T}\right)\left(T^\mathrm{T}-\bi b \bi a^\mathrm{T} \right).
\end{eqnarray}
The eigenvalues of $Q_A$ are the squares of the ellipsoid semiaxes $s_i$ and the eigenvectors give the orientation of these axes. Together with a specification of Bob's local basis, the geometric data $(\EA, \bi a, \bi b)$ provide a faithful representation of two-qubit states (Figure \ref{fig:generic_ellipsoid})~\cite{Jevtic2013}.

When Bob is steered by Alice, we can consider his ellipsoid ${\cal E}_B$, described by $\bi c_B$ and $Q_B$. This amounts to swapping $\bi a \leftrightarrow \bi b$ and $T \leftrightarrow T^\mathrm{T}$ in the expressions for $\bi c_A$ and $Q_A$.

Bob's steering of Alice is said to be \emph{complete} when, for any convex decomposition of $\bi{a}$ into states in $\EA$ or on its surface, there exists a POVM for Bob that steers to it~\cite{Jevtic2013}. All nonzero volume $\EA$ correspond to states that are completely steerable by Bob. When Bob's steering is complete, $\bi{a}$ lies on an ellipsoid $\EA$ scaled down by a factor $b=|\bi{b}|$; for incomplete steering of Alice, $\bi a$ lies strictly inside this scaled-down ellipsoid. Aside from these straightforward, necessary restrictions on $\bi{a}$ and $\bi{b}$, finding whether any two-qubit operator $\rho$ describes a physical state usually involves obscure functions of the components of the matrix $T$, resulting from the requirement that $\rho\geq 0$. However, these functions become much clearer in the context of the steering ellipsoid.

It will prove very useful to perform a reversible, trace-preserving local filtering operation that transforms $\rho$ to a canonical state $\widetilde{\rho}$. Crucially, Alice's steering ellipsoid is invariant under Bob's local filtering operation, so the same $\EA$ describes both $\rho$ and $\widetilde{\rho}$. We may perform the transformation~\cite{Shi2011}
\begin{eqnarray}
\label{eq:filtering}
\rho\rightarrow\widetilde \rho&=&\left(\mathbbm{1}\otimes \frac{1}{\sqrt{ 2\rho_B}}\right)\,\rho\,\left(\mathbbm{1}\otimes  \frac{1}{\sqrt{ 2\rho_B}}\right)\nonumber\\
&=&\frac{1}{4}(\mathbbm{1}\otimes\mathbbm{1}+\wt{\bi{ a}} \cdot\bsigma\otimes\mathbbm{1}+
\sum_{i,j=1}^3 \wt{T}_{ij}\,\sigma_i\otimes\sigma_j)
\end{eqnarray}
provided that Bob's reduced state $\rho_B=\tr_A \rho$ is invertible (the only exception occurs when $\rho_B$ is pure, in which case $\rho$ is a product state for which no steering is possible). In this canonical frame, Bob's state is maximally mixed ($\bi{\widetilde b}=\mathbf{0}$) and Alice's Bloch vector coincides with the centre of $\EA$ ($\bi{\widetilde a}=\bi{c}_A$). The ellipsoid matrix is given by $Q_A=\widetilde{T}\widetilde{T}^\mathrm{T}$, and so the semiaxes are $s_i=|t_i|$, where $t_i$ are the signed singular values of $\wt{T}$.

The local filtering operation preserves positivity: $\rho\geq 0$ if and only if $\widetilde \rho\geq 0$. It also maintains the separability of a state: $\rho$ is entangled if and only if $\widetilde \rho$ is~\cite{Verstraete2001}. We may therefore determine the positivity and separability of $\rho$ by studying its canonical state $\widetilde{\rho}$.

Applying state-dependent local unitary operations on $\widetilde{\rho}$, we can achieve the transformations $\bi{\widetilde a}\rightarrow O_A \bi{\widetilde a}$, $\bi{\widetilde b}\rightarrow O_B \bi{\widetilde b}$ and $\widetilde{T}\rightarrow O_A \widetilde{T} O_B^\mathrm{T}$ with $O_A,O_B\in\mathrm{SO(3)}$~\cite{Horodecki1996}. We can always find $O_A$ and $O_B$ that perform a signed singular value decomposition on $\wt T$, i.e. $O_A \widetilde{T} O_B^\mathrm{T} = \diag(\bi t)$. Bob's rotation $O_B$ has no effect on $\EA$, but $O_A$ rotates $\EA$ about the origin (treating $\bi{c}_A$ as a rigid rod) to align the semiaxes of $\EA$ parallel with the coordinate axes. Note there is some freedom in performing this rotation: the elements of $\bi{t}$ can be permuted and two signs can be flipped, but the product ${t}_1 {t}_2 {t}_3$ is fixed.

Both the positivity and entanglement of $\widetilde{\rho}$ are invariant under such local unitary operations. We therefore need only consider states that have $\EA$ aligned with the coordinates axes in this way. The question of physicality of \textit{any} general operator of the form \eref{eq:general_state} therefore reduces to considering canonical, aligned states
\begin{eqnarray}\label{eq:state}
\widetilde \rho=\frac{1}{4}(\mathbbm{1}\otimes\mathbbm{1}+\bi{c}_A\cdot\bsigma\otimes\mathbbm{1}+\sum_{i=1}^3{t}_i\,\sigma_i\otimes\sigma_i).
\end{eqnarray}
In the steering ellipsoid picture, this restricts our analysis to looking only at steering ellipsoids whose semiaxes are aligned with the coordinate axes: $Q_A=\diag(t_1^2, t_2^2, t_3^2)=\diag(s_1^2,s_2^2,s_3^2)$.

In the following, unless stated otherwise, we will only refer to Alice's steering ellipsoid; we therefore drop the label $A$ so that ${\E} \equiv {\EA}$, $Q \equiv Q_A$ and $\bi c \equiv \bi c_A$.

\section{Physical state conditions and chirality}\label{section_physical}

We now obtain conditions for the physicality of a two-qubit state $\widetilde \rho$ of the form \eref{eq:state}. The results of Braun et al.~\cite{Braun2013} employ Descartes' rules of signs to find when all the roots of the characteristic polynomial are non-negative; this shows that $\widetilde\rho\geq 0$ if and only if
\begin{eqnarray}
\det \widetilde\rho\geq 0 \text{ and } c^2\leq 1 - \sum_{i=1}^3{t}_i^2-2{t}_1 {t}_2 {t}_3 \text{ and } c^2 + \sum_{i=1}^3{t}_i^2 \leq 3.
\label{eq:braun_conds}
\end{eqnarray}

We find that $\det \widetilde \rho=\frac{1}{256}(c^4-2 u c^2 +q)$, where  $u = 1-\sum_{i} {t}_i^2+2\sum_{i} {t}_i^2 {\hat{c}_i}\!^2$, the unit vector $\bi{\hat c}=\bi{c}/c$ and 
\begin{equation}q = (1+{t}_1+{t}_2-{t}_3)(1+{t}_1-{t}_2+{t}_3)(1-{t}_1+{t}_2+{t}_3)(1-{t}_1-{t}_2-{t}_3).\end{equation}

To obtain geometric conditions for the physicality of $\wt\rho$, we express these conditions in terms of rotational invariants. Some care is needed with the term ${t}_1 {t}_2 {t}_3$, which could be positive or negative. Since $\sqrt{\det Q} = |t_1 t_2 t_3|=s_1 s_2 s_3 $ is positive by definition, we have that ${t}_1 {t}_2 {t}_3=\chi\sqrt{\det Q}$, where
\begin{eqnarray}
\chi=\mathrm{sign}(\det \wt T)= \mathrm{sign} ({t}_1 {t}_2 {t}_3)
\label{eq:w}
\end{eqnarray}
describes the \emph{chirality} of $\E$.

Let us say that Bob performs Pauli measurements on $\wt\rho$ and obtains the $+1$ eigenstates as outcomes, corresponding to Bloch vectors $\bi{\hat x}$, $\bi{\hat y}$ and $\bi{\hat z}$. These vectors form a right-handed set. These outcomes steer Alice to the Bloch vectors $\bi c+t_1\bi{\hat x}$, $\bi c+t_2\bi{\hat y}$ and $\bi c+t_3\bi{\hat z}$ respectively. When Bob's outcomes and Alice's steered vectors are related by an affine transformation involving a proper (improper) rotation, Alice's steered vectors form a right-handed (left-handed) set and $\chi=+1$ ($\chi=-1$). We therefore refer to $\chi=+1$ ellipsoids as \textit{right-handed} and $\chi=-1$ ellipsoids as \textit{left-handed}. Note that a degenerate ellipsoid corresponds to $\chi=0$, since at least one $t_i=0$ (equivalently $s_i = 0$).

\begin{theorem}\label{conditions}Let $\widetilde\rho$ be an operator of the form \eref{eq:state}, described by steering ellipsoid $\E$ with centre $\bi c$, matrix $Q$ and chirality $\chi$. This corresponds to a two-qubit state, $\widetilde\rho \geq 0$, if and only if
\begin{eqnarray*}
c^4-2 u c^2 +q\geq 0 \text{ and } c^2\leq 1 - \tr Q -2\chi\sqrt{\det Q} \text{ and } c^2 + \tr Q \leq 3,
\end{eqnarray*}where
\begin{eqnarray*}u& =1-\tr Q+2\bi{\hat c}^\mathrm{T}Q\bi{\hat c}, \\ q &=1+2\tr(Q^2)-2\tr Q - (\tr Q)^2-8 \chi \sqrt{\det Q}.\end{eqnarray*}

\begin{proof}
Rewrite the conditions in \eqref{eq:braun_conds} using the ellipsoid parameters $Q$, $\bi c$ and $\chi$.
\end{proof}
\end{theorem}

It should be noted that any $\E$ inside the Bloch sphere must obey $\sqrt{\det Q}\leq 1$. For such $\E$ we have that $c^2\leq 1 - \tr Q -2\chi\sqrt{\det Q} \Rightarrow c^2 + \tr Q \leq 3$ and hence the condition $c^2 + \tr Q \leq 3$ is redundant.

As with the criteria for entanglement given in equation (4) of Ref.~\cite{Jevtic2013}, we can identify three geometric contributions influencing whether or not a given steering ellipsoid describes a physical state: the distance of its centre from the origin, the size of the ellipsoid and the skew $\bi{\hat c}^\mathrm{T}Q\bi{\hat c}$. In addition, the physicality conditions also depend on the chirality of the ellipsoid, which relates to the separability of a state.

\begin{theorem}\label{chirality}Let $\widetilde\rho$ be a canonical two-qubit state of the form \eref{eq:state}, described by steering ellipsoid $\E$.
\begin{enumerate}
\item $\E$ for an entangled state $\widetilde\rho$ must be left-handed.
\item $\E$ for a separable state $\widetilde\rho$ may be right-handed, left-handed or degenerate. For a separable left-handed $\E$, the corresponding right-handed $\E$ is also a separable state and vice-versa.
\end{enumerate}
\newpage
\begin{proof}
$\,$
\begin{enumerate}
\item An entangled state $\widetilde\rho$ must have $\det \widetilde\rho^\mathrm{T_B} < 0$~\cite{Jevtic2013,detrhoref} (following from the Peres-Horodecki criterion) and a non-degenerate ellipsoid, hence $\chi=\pm 1$ \textit{a priori}. Partial transposition $\widetilde\rho\rightarrow\widetilde\rho\,^\mathrm{T_B}$ is equivalent to ${t}_2\rightarrow -{t}_2$ and hence to $\chi\rightarrow -\chi$. All quantum states achieve $\det\widetilde\rho\geq 0$, so for an entangled $\widetilde\rho$, we have $\det\widetilde\rho>\det \widetilde\rho^\mathrm{T_B}$. Using the form for $\det\widetilde\rho$ given in Theorem \ref{conditions}, an entangled canonical state must have $-8\chi\sqrt{\det Q}>8\chi \sqrt{\det Q}$ and so its chirality is restricted to $\chi=-1$.

\item The ellipsoid for a separable state may be degenerate or non-degenerate and so $\chi=0$ or $\chi=\pm 1$ \textit{a priori}. For a two-qubit separable state $\widetilde\rho$, the operator $\widetilde\rho^\mathrm{T_B}$ is also a separable state~\cite{Horodecki1996a}. Since partial transposition is equivalent to  $\chi\rightarrow -\chi$, this means that both the $\chi$ and the $-\chi$ ellipsoids are separable states. For the degenerate case, $\chi=0$. For a non-degenerate ellipsoid, both the $\chi=+1$ and $\chi=-1$ ellipsoids are separable states.
\end{enumerate}
\end{proof}
\end{theorem}

Recall that a local filtering transformation maintains the separability of a state. Although the chirality of an ellipsoid is a characteristic of canonical states only, we can extend Theorem \ref{chirality} to apply to any general state of the form \eref{eq:general_state} by defining the chirality of a general ellipsoid as that of its canonical state.

\begin{figure}
\begin{indented}
\item[]\includegraphics[width=0.6\textwidth]{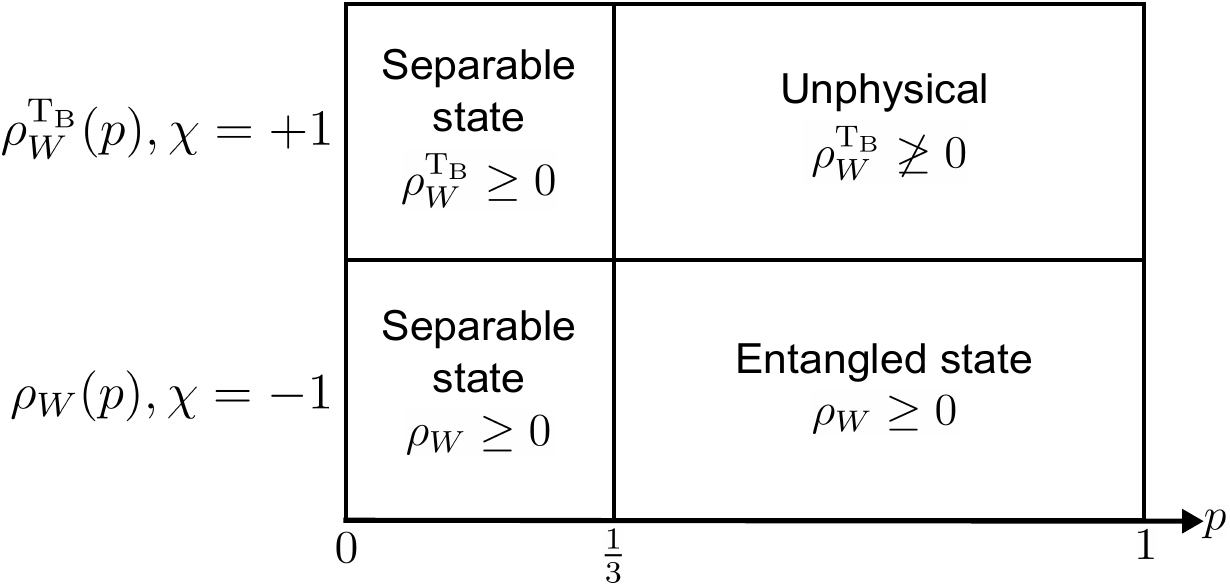}
\end{indented}
\caption{The physicality and separability of a steering ellipsoid depend on its chirality $\chi$. This dependence is illustrated for the set of Werner states $\rho_W(p)$ of the form \eqref{werner}.}
\label{fig:werner_state_chirality}
\end{figure}

As an example, consider the set of Werner states given by
\begin{eqnarray}\label{werner}
\rho_W(p)=p\ket{\psi^-}\bra{\psi^-}+\frac{1-p}{4}\mathbbm{1}\otimes\mathbbm{1},
\end{eqnarray}
where $\ket{\psi^-}=\frac{1}{\sqrt 2}(\ket{01}-\ket{10})$ and $0\leq p \leq 1$~\cite{Werner1989}. Although Werner's original definition does not impose the restriction $p\geq 0$, states with $p\leq0$ can be obtained from the partial transposition of states with $p\geq 0$. We will see in Section \ref{largest_spheres} that $\rho_W(p)$ is described by a spherical $\E$ of radius $p$ centred on the origin. Figure \ref{fig:werner_state_chirality} illustrates Theorem \ref{chirality} for $\rho_W(p)$.

\section{Extremal ellipsoid states}\label{section_maximal}

We will now use Theorems \ref{conditions} and \ref{chirality} to investigate ellipsoids lying on the entangled-separable and physical-unphysical boundaries by finding the largest area ellipses and largest volume ellipsoids with a given centre. The ellipsoid centre $\bi c$ is a natural parameter to use in the steering ellipsoid representation, and the physical and geometric results retrospectively confirm the relevance of this maximisation. In particular, we will see in Section \ref{section_concurrence} that the largest volume physical ellipsoids describe a set of states that maximise concurrence.

The methods used to find extremal ellipsoids are given in full in the Appendix, but the importance of Theorem \ref{chirality} should be highlighted. The ellipsoid of an entangled state must be left-handed. For non-degenerate $\E$ we can therefore probe the separable-entangled boundary by finding the set of extremal physical $\E$ with $\chi=+1$; these must correspond to extremal separable states. The physical-unphysical boundary is found by studying the set of extremal physical $\E$ with $\chi=-1$. Clearly the separable-entangled boundary must lie inside the physical-unphysical boundary since separable states are a subset of physical states. For the case of a degenerate $\E$ with $\chi=0$, any physical ellipsoid must be separable and so there is only the physical-unphysical boundary to find.

\subsection{Two dimensions: largest area circles and ellipses in the equatorial plane}\label{2d}

We begin by finding the physical-unphysical boundary for $\E$ lying in the equatorial plane. For a circle of radius $r$, centre $\bi c$, we find that $\E$ represents a physical (and necessarily separable) state if and only if $r\leq\frac{1}{2}(1-c^2)$.

The physical ellipse with the largest area in the equatorial plane is not a circle for $c > 0$. For $\E$ with centre $\bi c=(c, 0, 0)$, the maximal area physical ellipse has minor semiaxis $s_1=\frac{1}{4}(3-\sqrt{1+8c^2})$ and major semiaxis $s_2=\frac{1}{\sqrt{8}}\sqrt{1-4c^2+\sqrt{1+8c^2}}$. Noting that that both $s_1$ and $s_2$ are monotonically decreasing functions of $c$ with $s_1\leq s_2\leq \frac{1}{2}$, we see that the overall largest ellipse is the radius $\frac{1}{2}$ circle centred on the origin. Our results describe how a physical ellipse must shrink from this maximum as its centre is displaced towards the edge of the Bloch sphere.

Note that the unit disk does not represent a physical state; this corresponds to the well-known result that its Choi-isomorphic map is not CP (the `no pancake' theorem). In fact, Ref.~\cite{Braun2013} gives a generalisation of the no pancake theorem that immediately rules out such a steering ellipsoid: a physical steering ellipsoid can touch the Bloch sphere at a maximum of two points unless it is the whole Bloch sphere (as will be the case for a pure entangled two-qubit state).

\subsection{Three dimensions: largest volume spheres}\label{largest_spheres}

In three dimensions we find distinct separable-entangled and physical-unphysical boundaries. Inept states~\cite{Jones2005} form a family of states given by $\rho=r\ket{\phi_\epsilon}\bra{\phi_\epsilon}+(1-r)\rho'\otimes\rho'$, where $\ket{\phi_\epsilon}=\sqrt\epsilon\ket{00}+\sqrt{1-\epsilon}\ket{11}$ and $\rho'=\tr _A \ket{\phi_\epsilon}\bra{\phi_\epsilon}=\tr _B \ket{\phi_\epsilon}\bra{\phi_\epsilon}$. (The name `inept' was introduced because such states arise from the inept delivery of entangled qubits to pairs of customers: the supplier has a supply of pure entangled states $\ket{\phi_\epsilon}$ and always delivers a qubit to each customer but only has probability $r$ of sending the correct pair of qubits to any given pair of customers.) The two parameters $r$ and $\epsilon$ that describe an inept state can easily be translated into a description of the steering ellipsoid: $\E$ has $\bi c=(0, 0, (2\epsilon - 1)(1-r))$ and $Q=\mathrm{diag}(r^2, r^2, r^2)$. Thus an inept state gives a spherical $\E$ of radius $r$. Note that inept states with $\epsilon=\frac{1}{2}$ have null Bloch vectors for Alice and Bob and are equivalent to Werner states. The corresponding $\E$ are centred on the origin.

The separable-entangled boundary for a spherical $\E$ with centre $\bi c$ corresponds to $r=\frac{1}{3}(\sqrt{4-3c^2}-1)$. Any left- or right-handed sphere smaller than this bound describes a separable state. The physical-unphysical boundary is $r=1-c$. A spherical $\E$ on this boundary touches the edge of the Bloch sphere, and so this is just the constraint that $\E$ should lie inside the Bloch sphere. All left-handed spherical $\E$ inside the Bloch sphere therefore represent inept states. Right-handed spheres whose $r$ exceeds the separable-entangled bound cannot describe physical states since an entangled $\E$ must be left-handed. Note how simple the physical state criteria are for spherical $\E$: subject to these conditions on chirality, all spheres inside the Bloch sphere are physical. The same is not true for ellipsoids in general; there are some ellipsoids inside the Bloch sphere for which both the left- and right-handed forms are unphysical.

\subsection{Three dimensions: largest volume ellipsoids}

As explained in the Appendix, any maximal ellipsoid must have one of its axes aligned radially and the other two non-radial axes equal. The largest volume separable $\E$ centred at $\bi c$ is an oblate spheroid with its minor axis oriented radially. For an ellipsoid with $\bi c=(0, 0, c)$, the major semiaxes are $s_1=s_2=\frac{1}{\sqrt{18}}\sqrt{1-3c^2+\sqrt{1+3c^2}}$ and the minor semiaxis is $s_3=\frac{1}{3}(2-\sqrt{1+3c^2})$.

The largest volume physical $\E$ centred at $\bi c$ is also an oblate spheroid with its minor axis oriented radially. For an ellipsoid with $\bi c=(0, 0, c)$, the major semiaxes are $s_1=s_2=\sqrt{1-c}$ and the minor semiaxis is $s_3=1-c$. These extremal ellipsoids are in fact the largest volume ellipsoids with centre $\bi c$ that fit inside the Bloch sphere.

The volume $V$ of these maximal ellipsoids can be used as an indicator for entanglement and unphysicality. Our calculations have been carried out for a canonical $\widetilde\rho$, but since steering ellipsoids are invariant under the canonical transformation \ref{eq:filtering}, the results are directly applicable to any general $\rho$. The maximal ellipsoids for a general $\bi c$ are simply rotations of those found above for $\bi c=(0, 0, c)$; the results therefore depend only on the magnitude $c$.

\newpage

\begin{theorem}\label{indicator}Let $\rho$ be an operator of the form \eref{eq:general_state}, described by steering ellipsoid $\E$ with centre $\bi c$ and volume $V$. Let $V^{\mathrm{sep}}_c=\frac{2\pi}{81}\Big(1-9c^2+(1+3c^2)^{3/2}\Big)$ and $V^{\mathrm{max}}_c=\frac{4\pi}{3}(1-c)^2$.

\begin{enumerate}
\item If $\rho$ is a physical state and $V>V^{\mathrm{sep}}_c$ then $\rho$ must be entangled.
\item  If $V>V^{\mathrm{max}}_c$ then $\rho$ must be unphysical.
\end{enumerate}

\begin{proof}Find the volume of the ellipsoids on the separable-entangled and physical-unphysical boundaries using $V=\frac{4\pi}{3}s_1 s_2 s_3$.\end{proof}
\end{theorem}

This result extends the notion of using volume as an indicator for entanglement, as was introduced in Ref.~\cite{Jevtic2013}. We see that the largest volume separable ellipsoid is the Werner state on the separable-entangled boundary, which has a spherical $\E$ of radius $\frac{1}{3}$ and $c=0$. We have tightened the bound by introducing the dependence on $c$. In fact, Theorem \ref{indicator} gives the tightest possible such bounds, since we have identified the extremal $\E$ that lie on the boundaries. Note that for all $c$ we have $V^{\mathrm{sep}}_c \leq V^{\mathrm{max}}_c$, with equality achieved only for $c=1$ when $\E$ is a point with $V=0$ and $\rho$ is a product state. This confirms that the two boundaries are indeed distinct and that the separable $\E$ are a subset of physical $\E$.

\subsection{Applications to classical Euclidean geometry using the nested tetrahedron condition}\label{section_geometry}

Recall the nested tetrahedron condition~\cite{Jevtic2013}: a two-qubit state is separable if and only if $\E$ fits inside a tetrahedron that fits inside the Bloch sphere. We used Theorem \ref{conditions} and ellipsoid chirality to algebraically find the separable-entangled boundary for the cases that $\E$ is a circle, ellipse, sphere or ellipsoid. The nested tetrahedron condition then allows us to derive several interesting results in classical Euclidean geometry. We give a very brief summary of the work here; a full discussion is given in Ref.~\cite{GeometryPaper}.

Euler's inequality $r\leq \frac{R}{2}$ is a classic result relating a triangle's circumradius $R$ and inradius $r$~\cite{RecentAdvances}. In Section \ref{2d} we investigated the largest circular $\E$ in the equatorial plane, finding that $\E$ represented a physical (and necessarily separable) state if and only if $r\leq\frac{1}{2}(1-c^2)$. By the degenerate version of the nested tetrahedron condition, this gives the condition for when $\E$ fits inside a triangle inside the unit disk ($R=1$). We therefore see that our result implies Euler's inequality, since $0 \leq c \leq 1$.

We can pose the analogous question in 3 dimensions. Let ${\cal{S}}_r$ be a sphere of radius $r$ contained inside another sphere ${\cal{S}}_R$ of radius $R$. If the distance between the sphere centres is $c$, what are the necessary and sufficient conditions for the existence of a tetrahedron circumscribed about ${\cal{S}}_r$ and inscribed in ${\cal{S}}_R$? This question was answered by Danielsson using some intricate projective geometry~\cite{Danielsson}, but there is no known proof using only methods belonging to classical Euclidean elementary geometry~\cite{Enigma}. By considering the steering ellipsoids of inept states (Section \ref{largest_spheres}) we have answered precisely this question, finding the necessary and sufficient condition for the existence a nested tetrahedron. Our result is found to reproduce Danielsson's result that the sole condition is $c^2\leq (R+r)(R-3r)$.

In fact, our work extends these results to give conditions for the existence of a nested tetrahedron for the more general case of an ellipsoid $\E$ contained inside a sphere. These very non-trivial geometric results can be straightforwardly derived from Theorem \ref{conditions} by understanding the separability of two-qubit states in the steering ellipsoid formalism.

\section{Applications to mixed state entanglement: ellipsoid volume and concurrence}\label{section_concurrence}

The volume of a state provides a measure of the quantum correlations between Alice and Bob, distinct from both entanglement and discord~\cite{Jevtic2013}. We will now study the states corresponding to the maximal volume physical ellipsoids. By deriving a bound for concurrence in terms of ellipsoid volume, we see that maximal volume states also maximise concurrence for a given ellipsoid centre.

\subsection{Maximal volume states}

\begin{figure}
\begin{indented}
\item[]\includegraphics[width=0.85\textwidth]{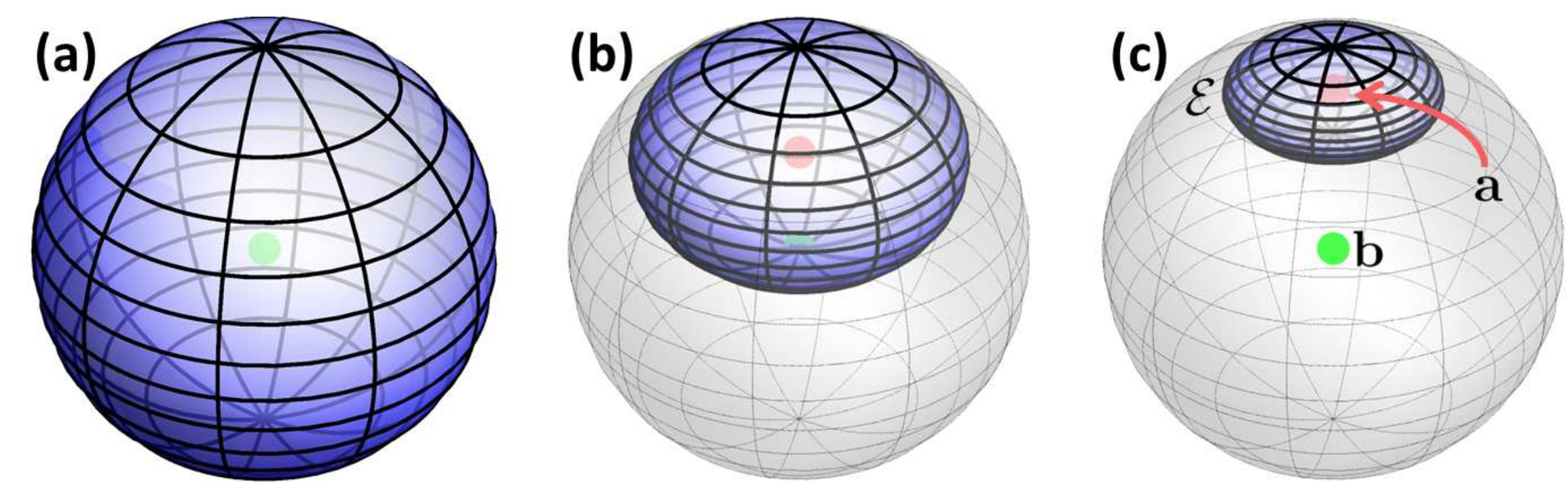}
\end{indented}
\caption{The geometric data for three maximal volume states $\widetilde\rho^\mathrm{\,max}_c$ with (a) $c=0$, (b) $c=0.5$ and (c) $c=0.8$. Since these are canonical states we have $\bi b=\mathbf{0}$ and $\bi c = \bi a$. Note that ${\E}={\E}^{\max}_c$ touches the North pole of the Bloch sphere for any $c$.}
\label{fig:maximal_volume_ellipsoids}
\end{figure}

Recall that the largest volume ellipsoid with $\bi c=(0, 0, c)$ has major semiaxes $s_1=s_2=\sqrt{1-c}$ and minor semiaxis $s_3=1-c$. We will call this ${\E}^{\max}_c$. With the exception of $c=1$, which describes a product state, these correspond to entangled states and so are described by left-handed steering ellipsoids. Using \eref{eq:state}, the canonical state for ${\E}^{\max}_c$ is
\begin{eqnarray}\label{eq:max_state}
\widetilde\rho^\mathrm{\,max}_c=\left(1-\frac{c}{2}\right)\ket{\psi_{c}}\bra{\psi_{c}}+\frac{c}{2}\ket{00}\bra{00},
\end{eqnarray}
where $\ket{\psi_{c}}=\frac{1}{\sqrt{2-c}}(\ket{01}+\sqrt{1-c}\ket{10})$.
This describes a family of rank-2 `X states' parametrised by $0\leq c\leq 1$. Some examples are shown in Figure \ref{fig:maximal_volume_ellipsoids}.

The density matrix of an X state in the computational basis has non-zero elements only on the diagonal and anti-diagonal, giving it a characteristic X shape. X states were introduced in Ref.~\cite{Yu2007} as they comprise a large class of two-qubit states for which certain correlation properties can be found analytically. In fact, steering ellipsoids have already been used to study the quantum discord of X states~\cite{Shi2011}. In the steering ellipsoid formalism, $\E$ for an X state will be radially aligned, having a semiaxis collinear with $\bi c$.

The state $\widetilde\rho^\mathrm{\,max}_c$ should be compared to the Horodecki state $\rho^\mathrm{H}=p\ket{\psi^+}\bra{\psi^+}+(1-p)\ket{00}\bra{00}$, where $\ket{\psi^+}=\frac{1}{\sqrt 2}(\ket{01}+\ket{10})$; this is the same as $\widetilde\rho^\mathrm{\,max}_c$ when we reparametrise $c=2(1-p)$ and also make the change $\ket{\psi^+}\rightarrow\ket{\psi_{c}}$. The Horodecki state is a rank-2 maximally entangled mixed state~\cite{Ishizaka2000}. $\rho^\mathrm{H}$ may be extended (see, for example, Refs.~\cite{Horst2013,Bartkiewicz2,Miranowicz2008}) to the generalised Horodecki state $\rho^\mathrm{GH}=p\ket{\psi_\alpha}\bra{\psi_\alpha}+(1-p)\ket{00}\bra{00}$, where $\ket{\psi_\alpha}=\sqrt{\alpha}\ket{01}+\sqrt{1-\alpha}\ket{10}$. Note that this has two free parameters, $\alpha$ and $p$. Setting $\alpha=1/2p$ and reparametrising $c=2(1-p)$, we see that our $\widetilde\rho^\mathrm{\,max}_c$ states form a special class of the generalised Horodecki states described by the single parameter $c$.

The maximal volume states have a clear physical interpretation when we consider the Choi-isomorphic channel: $\widetilde\rho^\mathrm{\,max}_c$ is isomorphic to the single qubit amplitude-damping (AD) channel with decay probability $c$~\cite{Horst2013}. For a single qubit state $\eta$, this channel is $\Phi_\mathrm{AD}(\eta)=E_0\eta E_0^\dagger+E_1\eta E_1^\dagger$, where~\cite{Nielsen2010}\begin{eqnarray}\nonumber E_0=\begin{pmatrix}1 & 0 \\ 0 & \sqrt{1-c} \end{pmatrix} \text { and }\, E_1=\begin{pmatrix}0 & \sqrt{c} \\ 0 & 0 \end{pmatrix}.\end{eqnarray}

If Alice and Bob share the Bell state $\ket{\psi^+}=\frac{1}{\sqrt{2}}(\ket{01}+\ket{10})$ and Alice passes her qubit through this channel, we obtain a maximal volume state centred at $\bi c=(0,0,c)$, i.e. $\widetilde\rho^\mathrm{\,max}_c=(\Phi_\mathrm{AD}\otimes\mathbbm{1})(\ket{\psi^+}\bra{\psi^+})$.

\subsection{Bounding concurrence using ellipsoid volume}

Physically motivated by its connection to the entanglement of formation~\cite{Wootters1998}, concurrence is an entanglement monotone that may be easily calculated for a two-qubit state $\rho$. Define the spin-flipped state as $\hat{\rho}=(\sigma_y \otimes \sigma_y)\rho^*(\sigma_y \otimes \sigma_y)$ and let $\lambda_1,...,\lambda_4$ be the square roots of the eigenvalues of $\rho \hat\rho$ in non-increasing order. The concurrence is then given by
\begin{eqnarray}
C(\rho)=\max(0, \lambda_1-\lambda_2-\lambda_3-\lambda_4).
\end{eqnarray}

Concurrence ranges from 0 for a separable state to 1 for a maximally entangled state. In principle one may find $C(\rho)$ in terms of the parameters describing the corresponding steering ellipsoid $\E$, but the resulting expressions are very complicated. It is however possible to derive a simple bound for $C(\rho)$ in terms of steering ellipsoid volume.

\begin{lemma}\label{t_state_concurrence}Let $\tau$ be a Bell-diagonal state given by \begin{eqnarray}\label{eq:t_state}\tau=\frac{1}{4}(\mathbbm{1}\otimes\mathbbm{1}+\sum_{i=1}^3 t_{i}\,\sigma_i\otimes\sigma_i).\end{eqnarray} The concurrence is bounded by $C(\tau)\leq \sqrt{|t_1 t_2 t_3|}$, and there exists a state $\tau$ that saturates the bound for any value $0\leq C(\tau)\leq 1$.
\begin{proof}
Without loss of generality, order $t_1\geq t_2 \geq |t_3|$. Ref.~\cite{Verstraete2001} then gives $C(\tau)=\max\{0,\frac{1}{2}(t_1+t_2-t_3-1)\}$. For a separable state $\tau$, we have $C(\tau)=0$ and so the bound holds.

An entangled state $\tau$ must have $C(\tau)>0$. Recalling that the semiaxes $s_i=|t_i|$ and that an entangled state must have $\chi=-1$ (Theorem \ref{chirality}), we take $t_1=s_1$, $t_2=s_2$ and $t_3=-s_3$ to obtain $C(\tau)=\frac{1}{2}(s_1+s_2+s_3-1)$.

Ref.~\cite{Horodecki1996} gives necessary and sufficient conditions for the positivity and separability of $\tau$. For $\tau$ to be an entangled state, the vector $\bi{s}=(s_1,s_2,s_3)$ must lie inside the tetrahedron with vertices $\bi{r}_0=(1,1,1)$, $\bi{r}_1=(1,0,0)$, $\bi{r}_2=(0,1,0)$ and $\bi{r}_3=(0,0,1)$. Since the tetrahedron $(\bi r_0, \bi r_1, \bi r_2, \bi r_3)$ is a simplex, we may uniquely decompose any point inside it as $\bi s=p_0 \bi r_0+p_1 \bi r_1+p_2 \bi r_2+p_3 \bi r_3$ where $\sum_i p_i=1$ and $0\leq p_i \leq 1$. This gives $\bi s=(p_0+p_1,p_0+p_2,p_0+p_3)$. Evaluating $s_1+s_2+s_3$, we obtain $C(\tau)=p_0$, as $\sum_i p_i=1$.

Now we evaluate the right hand side of the inequality. We have $|t_1 t_2 t_3|=s_1 s_2 s_3=(p_0+p_1)(p_0+p_2)(p_0+p_3)=p_0^2+p_0(p_1 p_2 + p_2 p_3 + p_3 p_1)+p_1 p_2 p_3$, where we have again used $\sum_i p_i=1$. Since all the terms are positive, we see that $\sqrt{|t_1 t_2 t_3|}\geq p_0=C(\tau)$, as required. The bound is saturated by states whose $\bi s$ vectors lie on the edges of the tetrahedron $(\bi r_0, \bi r_1, \bi r_2, \bi r_3)$. For example, by choosing $p_1=p_2=0$, we obtain the set of states $\bi s=(p_0, p_0, 1)$. These saturate the bound for any value of the parameter $0\leq p_0\leq 1$.
\end{proof}
\end{lemma}

\begin{theorem}\label{concurrence_vol}Let $\rho$ be a general two-qubit state of the form \eqref{eq:general_state}. The concurrence is bounded by $C(\rho)\leq \gamma_b^{-1}\big(\frac{3V}{4\pi}\big)^{1/4}$, where the Lorentz factor $\gamma_b=1/\sqrt{1-b^2}$ and $V$ is the volume of Alice's steering ellipsoid $\E$.
\begin{proof}

Any state $\rho$ can be transformed into a Bell-diagonal state $\tau$ of the form \eqref{eq:t_state} by local filtering operations~\cite{Verstraete2001}: $\tau = (A \otimes B)\rho(A\otimes B)^\dag / N$, where the normalisation factor $N=\tr((A \otimes B)\rho(A\otimes B)^\dag)$. The concurrence transforms as $C(\tau)=C(\rho)|\det A||\det B| / N$. Express the state $\rho$ in the Pauli basis $\{\mathbbm{1}, \bsigma\}^{\otimes 2}$ using the matrix $\Theta(\rho)$ whose elements are defined by $[\Theta(\rho)]_{\mu \nu}=\tr(\rho\, \sigma_\mu \otimes \sigma_\nu)$. Similarly $\tau$ is represented in the Pauli basis by $\Theta(\tau)$. The local filtering operations achieve $\Theta(\tau)=L_A \Theta(\rho) L_B^\mathrm{T} |\det A||\det B| / N$, where $L_A$ and $L_B$ are proper orthochronous Lorentz transformations given by $L_A = \Upsilon (A \otimes A^*)\Upsilon^{\dag} / |\det A|$, $L_B = \Upsilon (B \otimes B^*)\Upsilon^{\dag} / |\det B|$ with\begin{eqnarray}\nonumber\Upsilon=\frac{1}{\sqrt{2}}\begin{pmatrix}1 & 0 & 0 & 1 \\ 0 & 1 & 1 & 0 \\ 0 & i & -i & 0 \\ 1 & 0 & 0 & -1\end{pmatrix}.\end{eqnarray}

For a general state $\rho$, the volume $V=\frac{4\pi}{3}\gamma_b^4 |\det \Theta(\rho)|$~\cite{Jevtic2013}. From the local filtering transformation, and using $\det L_A=\det L_B=1$, we have\begin{eqnarray}\nonumber|\det \Theta(\tau)|=|\det \Theta(\rho)|\left(\frac{|\det A||\det B|}{N}\right)^4=|\det \Theta(\rho)|\left(\frac{C(\tau)}{C(\rho)}\right)^4.\end{eqnarray}
For a Bell-diagonal state $|\det \Theta(\tau)|=|t_1 t_2 t_3|$ and so we obtain $V=\frac{4\pi}{3}\gamma_b^4 |t_1 t_2 t_3|\left(\frac{C(\rho)}{C(\tau)}\right)^4$ and hence $C(\tau)=\left(\frac{4\pi}{3V}\right)^{1/4}\gamma_b |t_1 t_2 t_3|^{1/4} C(\rho)$. Since $|t_1 t_2 t_3|\leq 1$, Lemma \ref{t_state_concurrence} implies that $C(\tau)\leq |t_1 t_2 t_3|^{1/4}$, from which the result then follows.\end{proof}
\end{theorem}

This bound will be of central importance in the derivation of the CKW inequality in Section \ref{section_monogamy}. Theorem \ref{concurrence_vol} also suggests how ellipsoid volume might be interpreted as a quantum correlation feature called \emph{obesity}. If we define the obesity of a two-qubit state as $\Omega(\rho)=|\det \Theta(\rho)|^{1/4}$ then Theorem \ref{concurrence_vol} shows that concurrence is bounded for any two-qubit state as $C(\rho)\leq\Omega(\rho)$. Note that this definition also suggests an obvious generalisation to a $d$-dimensional Hilbert space, $\Omega(\rho)=|\det \Theta(\rho)|^{1/d}$.

\subsection{Maximal volume states maximise concurrence}

We now demonstrate the physical significance of the maximum volume steering ellipsoids by finding that the corresponding states $\widetilde\rho^\mathrm{\,max}_c$ maximise concurrence for a given ellipsoid centre. This will also demonstrate the tightness of the bound given in Theorem \ref{concurrence_vol}.

The state $\widetilde\rho^\mathrm{\,max}_c$ given in \eqref{eq:max_state} is a canonical state with $\bi{\widetilde b}=\mathbf{0}$. Let us invert the transformation \eqref{eq:filtering} to convert $\widetilde\rho^\mathrm{\,max}_c$ to a state with $\bi b \neq \mathbf{0}$: 
\begin{eqnarray}
\label{eq:rhomax_gen}
\widetilde\rho^\mathrm{\,max}_c\rightarrow \rho^\mathrm{\max}_c=&\left(\mathbbm{1}\otimes \sqrt{ 2\rho_B}\right)\,\widetilde\rho^\mathrm{\,max}_c\,\left(\mathbbm{1}\otimes\sqrt{ 2\rho_B}\right).\end{eqnarray}
This alters Bob's Bloch vector to $\bi b$, where $\rho_B=\frac{1}{2}(\mathbbm{1}+\bi b \cdot \bsigma)$ is Bob's reduced state. Recall that Bob's local filtering operation leaves Alice's steering ellipsoid $\E$ invariant, and so $\E$ for $\rho^\mathrm{\max}_c$ is still the maximal volume ellipsoid ${\E}^\mathrm{\max}_c$.

\begin{theorem}\label{max_vol_conc}
From the set of all two-qubit states that have $\E$ centred at $\bi c$, the state with the highest concurrence is $\widetilde\rho^\mathrm{\,max}_c$, as given in \eqref{eq:max_state}. The bound of Theorem \ref{concurrence_vol} is saturated for any $0 \leq b \leq 1$ by states $\rho^\mathrm{\,max}_c$ of the form \eqref{eq:rhomax_gen}, corresponding to the maximal volume ellipsoid ${\E}^{\max}_c$.
\begin{proof}
Recall that under the local filtering operation $\rho \rightarrow (A \otimes B)\rho(A\otimes B)^\dag / N$ concurrence transforms as $C(\rho)\rightarrow C(\rho)|\det A||\det B| / N$, where $N=\tr((A \otimes B)\rho(A\otimes B)^\dag)$~\cite{Verstraete2001}. For the canonical transformation \eqref{eq:filtering}, we have $A=\mathbbm{1}$ and $B=1/\sqrt{2\rho_B}$. This gives $\det A=1$, $\det B=\gamma_b$, $N=1$ so that $C(\wt\rho)=\gamma_b C(\rho)$. Computing the concurrence of \eqref{eq:max_state} gives $C(\wt\rho^\mathrm{\,max}_c)=\sqrt{1-c}$. Hence for a state of the form \eqref{eq:rhomax_gen} we have $C(\rho^\mathrm{\,max}_c)=\gamma_b^{-1} \sqrt{1-c}$.

Since $\E$ is invariant under Bob's local filtering operation, the same ${\E}^{\max}_c$ describes a state $\rho^\mathrm{\,max}_c$ with any $\bi b$. From Theorem \ref{indicator} we know that the maximal ellipsoid ${\E}^{\max}_c$ has volume $V^{\max}_c=\frac{4\pi}{3}(1-c)^2$. Substituting $C(\rho^\mathrm{\,max}_c)$ and $V^{\max}_c$ into the bound of Theorem \ref{concurrence_vol} shows that the bound is saturated by states $\rho^\mathrm{\,max}_c$ for any $0\leq b\leq 1$.

Any physical $\rho$ with $\E$ centred at $\bi c$ must obey the bounds $V\leq \frac{4\pi}{3}(1-c)^2$ (Theorem \ref{indicator}) and $C(\rho)\leq \gamma_b^{-1}\big(\frac{3V}{4\pi}\big)^{1/4}$ (Theorem \ref{concurrence_vol}), and hence $C(\rho)\leq \gamma_b^{-1} \sqrt{1-c}$. For a given $c$, the state that maximises concurrence has $b=0$. The state $\widetilde\rho^\mathrm{\,max}_c$ then achieves this maximum possible concurrence, $C(\widetilde\rho^\mathrm{\,max}_c)=\sqrt{1-c}$. Hence, from the set of all two-qubit states that have $\E$ centred at $\bi c$, the state with the highest concurrence is $\widetilde\rho^\mathrm{\,max}_c$.
\end{proof}
\end{theorem}

Note that $\widetilde\rho^\mathrm{\,max}_c$ maximises obesity from the set of all two-qubit states that have $\E$ centred at $\bi c$, achieving $\Omega(\widetilde\rho^\mathrm{\,max}_c)=\sqrt{1-c}$. Although the maximal volume ${\E}^\mathrm{\max}_c$ describes states $\rho^\mathrm{\,max}_c$ with any $\bi b$, the maximally obese state is uniquely the canonical $\widetilde\rho^\mathrm{\,max}_c$. The family of maximally obese states is studied further in~\cite{ObesityPaper}, with $\widetilde\rho^\mathrm{\,max}_c$ found to maximise several measures of quantum correlation in addition to concurrence.

\section{Monogamy of steering}\label{section_monogamy}

The maximal volume states $\rho^\mathrm{\,max}_c$ have particular significance when studying a monogamy scenario involving three qubits. Monogamy scenarios and steering ellipsoids have been used before to study the Koashi-Winter relation~\cite{Shi2011}. Here we show that ellipsoid volume obeys a monogamy relation that is strictly stronger than the CKW inequality for concurrence monogamy, giving us a new derivation of the CKW result. Subscripts labelling the qubits $A$, $B$ and $C$ are reintroduced so that Alice's ellipsoid $\E$ is now called $\EA$, the maximal volume state $\rho^\mathrm{\,max}_c$ is now called $\rho^\mathrm{\,max}_{c_A}$, and so on.

We begin by considering a maximal volume two-qubit state shared between Alice and Bob.

\begin{lemma}\label{mutual}If Alice and Bob share a state $\rho^\mathrm{max}_{c_A}$ given by \eqref{eq:rhomax_gen} then both $\EA$ and $\EB$ are maximal volume for their respective centres $\bi c_A$ and $\bi c_B$. The steering ellipsoid centres obey $\gamma_b^2 (1-c_B)=\gamma_a^2 (1-c_A)$.
\begin{proof}
$\EA={\E}^{\max}_{c_A}$ by construction, so $V_A=V_{c_A}^\mathrm{max}=\frac{4\pi}{3}(1-c_A)^2$. From Theorem \ref{max_vol_conc} we know that $C(\rho^\mathrm{\,max}_{c_A})=\gamma_b^{-1} \sqrt{1-c_A}$. Since concurrence is a symmetric function with respect to swapping Alice and Bob we must also have $C(\rho^\mathrm{\,max}_{c_A})=\gamma_a^{-1} \sqrt{1-c_B}$, which gives $\gamma_b^2 (1-c_B)=\gamma_a^2 (1-c_A)$. For any two-qubit state, the volumes of $\EA$ and $\EB$ are related by $\gamma_b^4 V_B = \gamma_a^4 V_A$~\cite{Jevtic2013}, so $V_B=\frac{4\pi}{3}(1-c_B)^2$. This means that $V_B=V_{c_B}^{\max}$ and so $\EB$ is also maximal volume for the centre $\bi c_B$, i.e. $\EB={\E}_{c_B}^{\max}$.
\end{proof}
\end{lemma}

\begin{figure}
\begin{indented}
\item[]\includegraphics[width=0.85\textwidth]{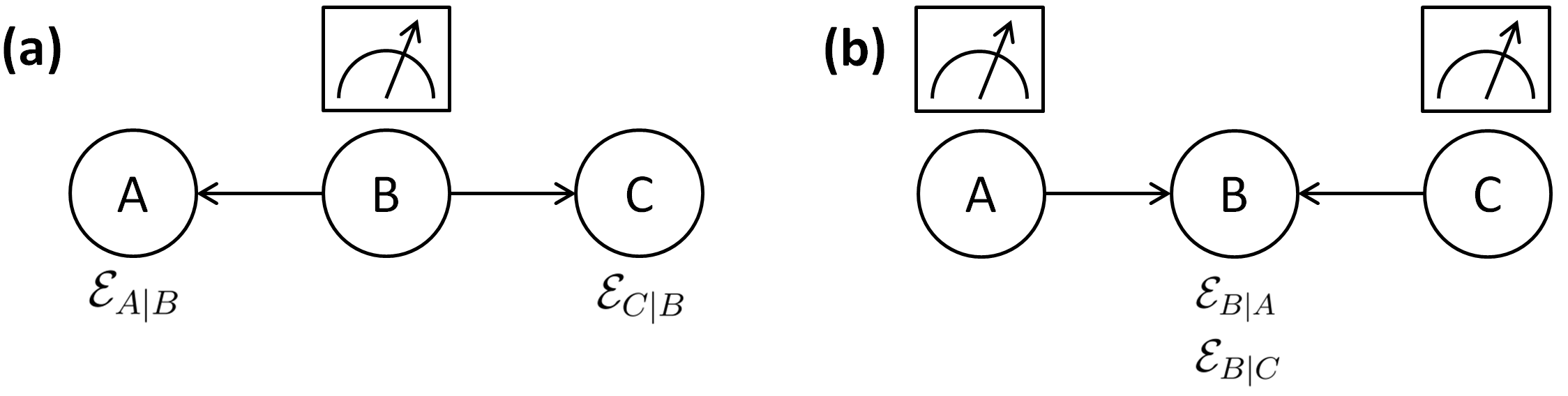}
\end{indented}
\caption{The two scenarios for studying monogamy, with arrows between parties indicating the direction of steering. (a) Bob performs a measurement to steer Alice and Charlie, with corresponding steering ellipsoids $\EAB$ and $\ECB$ respectively. (b) Alice and Charlie perform measurements to steer Bob, with corresponding steering ellipsoids $\EBA$ and $\EBC$ respectively.}
\label{fig:monogamy}
\end{figure}

Now consider Scenario (a) shown in Figure \ref{fig:monogamy}, in which Alice, Bob and Charlie share a pure three-qubit state and Bob can perform a measurement to steer Alice and Charlie. Let $\EAB$, with volume $V_{A|B}$ and centre $\bi c_{A|B}$, be the ellipsoid for Bob steering Alice, and similarly for the ellipsoid $\ECB$ with Bob steering Charlie.

\begin{lemma}\label{monogamy}Alice, Bob and Charlie share a pure three-qubit state for which the joint state $\rho_{AB}=\rho^\mathrm{max}_{c_{A|B}}$ given by $\eqref{eq:rhomax_gen}$, corresponding to $\EAB$ being maximal volume. The ellipsoid $\ECB$ is then also maximal volume, and the centres obey $c_{A|B}+c_{C|B}=1$.
\begin{proof}Consider first the case that Alice and Bob's state is the canonical $\rho_{AB}=\widetilde\rho^\mathrm{\,max}_{c_{A|B}}$ given by \eref{eq:max_state}, which means that $\EAB={\E}^{\max}_{c_{A|B}}$ by construction. Call the pure three-qubit state $\ket{\widetilde\phi_{ABC}}$, so that $\rho_{AB}=\tr_{C} \ket{\widetilde\phi_{ABC}} \bra{\widetilde\phi_{ABC}}$, with Bob's local state being maximally mixed. Performing a purification over Charlie's qubit, we obtain the rank-2 state $\ket{\widetilde\phi_{ABC}} = \frac{1}{\sqrt 2}(\sqrt{c_{A|B}}\ket{001}+\ket{010}+\sqrt{1-c_{A|B}}\ket{100})$. Finding $\rho_{BC}=\tr_{A} \ket{\widetilde\phi_{ABC}} \bra{\widetilde\phi_{ABC}}$, we see that the state $\rho_{BC}$ corresponds to a maximal volume $\ECB={\E}^{\max}_{c_{C|B}}$ with centre $c_{C|B}=1-c_{A|B}$.

Transforming out the the canonical frame $\ket{\widetilde\phi_{ABC}}\rightarrow\ket{\phi_{ABC}}=(\mathbbm{1}\otimes\sqrt{2\rho_B}\otimes \mathbbm{1})\ket{\widetilde\phi_{ABC}}$ `boosts' Bob's Bloch vector to an arbitrary $\bi b$, but leaves both $\EAB$ and $\ECB$ invariant. Therefore the relationship $c_{A|B}+c_{C|B}=1$ must also hold for the general case that $\rho_{AB}=\rho^\mathrm{max}_{c_{A|B}}$ with any $\bi b$.
\end{proof}
\end{lemma}

We now derive two monogamy relations for ellipsoid volume. The first relation concerns Scenario (a) discussed above, in which Bob can perform a measurement to steer Alice and Charlie. We are interested in the relationship between $V_{A|B}$ and $V_{C|B}$: does Bob's steering of Alice limit the extent to which he can steer Charlie? The second relation concerns Scenario (b) shown in Figure \ref{fig:monogamy}, in which Alice and Charlie can perform local measurements to steer Bob. We label the corresponding steering ellipsoids $\EBA$ and $\EBC$ respectively.

\begin{theorem}$\,$\label{volume_bounds}
\begin{enumerate}
\item When Alice, Bob and Charlie share a pure three-qubit state the ellipsoids steered by Bob must obey the bound $\sqrt{V_{A|B}}+\sqrt{V_{C|B}}\leq\sqrt{\frac{4\pi}{3}}$. The bound is saturated when $\EAB$ and $\ECB$ are maximal volume.  
\item When Alice, Bob and Charlie share a pure three-qubit state the ellipsoids steered by Alice and Charlie must obey the bound $\gamma_a^{-2}\sqrt{V_{B|A}}+\gamma_c^{-2}\sqrt{V_{B|C}}\leq\gamma_b^{-2}\sqrt{\frac{4\pi}{3}}$. The bound is saturated when $\EBA$ and $\EBC$ are maximal volume. 
\end{enumerate}

\begin{proof}\footnote{The original proof published in Ref.~\cite{Monogamy} is incorrect; a corrected version appears in Ref.~\cite{Corrigendum} and has been incorporated here. We are very grateful to Michael Hall for his assistance with the correction. In fact, the corrected proof reveals a remarkable new result relating the volume of Alice's steering ellipsoid to the centre of Charlie's: $V_{A|B}=\frac{4\pi}{3}c_{C|B}^2$. This implies that there is also a monogamy relation for steering ellipsoid centre: $c_{A|B}+c_{C|B}\leq 1$ (assuming that $\rho_B$ is non-singular so that Bob can steer). Note that Lemma~\ref{monogamy} manifestly follows from these new results: $\EAB$ is maximal volume ($V_{A|B}=V_{c_{A|B}}^\mathrm{max}=\frac{4\pi}{3}(1-c_{A|B})^2=\frac{4\pi}{3}c_{C|B}^2$) if and only if $\ECB$ is maximal volume ($V_{C|B}=V_{c_{C|B}}^\mathrm{max}=\frac{4\pi}{3}(1-c_{C|B})^2=\frac{4\pi}{3}c_{A|B}^2$), with $c_{A|B}+c_{C|B}=1$ holding.}
$\,$
\begin{myenumerate}
\item{Alice, Bob and Charlie hold the pure three-qubit state $\ket{\phi_{ABC}}$. The canonical transformation $\ket{\phi_{ABC}}\rightarrow\ket{\widetilde\phi_{ABC}}=(\mathbbm{1}\otimes\frac{1}{\sqrt{2\rho_B}}\otimes \mathbbm{1})\ket{\phi_{ABC}}$ leaves $\EAB$ and $\ECB$ invariant. We therefore need consider only canonical states for which $\bi{\wt b}=\mathbf{0}$. (When $\rho_B$ is singular and the canonical transformation cannot be performed, no steering by Bob is possible; we then have $V_{A|B}=V_{C|B}=0$ so that the bound holds trivially.)

We begin by showing that $V_{A|B}=\frac{4\pi}{3}c_{C|B}^2$. Denote the eigenvalues of $\wt{\rho}_{AB}=\tr_{C} \ket{\widetilde\phi_{ABC}} \bra{\widetilde\phi_{ABC}}$ as $\{\lambda_i\}$. For a canonical state Charlie's Bloch vector coincides with $\bi c_{C|B}$, and so $\wt\rho_{C}=\tr_{AB} \ket{\widetilde\phi_{ABC}} \bra{\widetilde\phi_{ABC}}=\frac{1}{2}(\mathbbm{1}+\bi c_{C|B}\cdot\bsigma)$. By Schmidt decomposition we therefore have $\{\lambda_i\}=\{\frac{1}{2}(1+c_{C|B}), \frac{1}{2}(1-c_{C|B}), 0, 0\}$. From the expression for $V_{A|B}$ given in Ref.~\cite{Jevtic2013} we obtain $V_{A|B}=\frac{64\pi}{3}|\det \wt\rho_{AB}^\mathrm{T_A}|$. Define the reduction map~\cite{Reduction1, Reduction2} as $\Lambda(X)=\mathbbm{1}\mathrm{tr}X-X$. Following Ref.~\cite{HorodeckiReduction} we note that $ \det \wt\rho_{AB}^\mathrm{T_A}=\det\,((\sigma_y\otimes\I)\wt\rho_{AB}^\mathrm{T_A}(\sigma_y\otimes\I))$ and that $(\sigma_y\otimes\I)\wt\rho_{AB}^\mathrm{T_A}(\sigma_y\otimes\I)=(\Lambda\otimes\mathbbm{1})(\wt\rho_{AB})=\frac{1}{2}\mathbbm{1}\otimes\mathbbm{1}-\wt\rho_{AB}$, where we have used the fact that Bob's local state is maximally mixed. Since the eigenvalues of $\frac{1}{2}\mathbbm{1}\otimes\mathbbm{1}-\wt\rho_{AB}$ are $\{\frac{1}{2}-\lambda_i\}$ we obtain $\det \wt\rho_{AB}^\mathrm{T_A}=\prod_i (\frac{1}{2}-\lambda_i)=(-\frac{1}{2}c_{C|B})(\frac{1}{2}c_{C|B})(\frac{1}{2})(\frac{1}{2})=-\frac{1}{16}c_{C|B}^2$, which gives $V_{A|B}=\frac{4\pi}{3}c_{C|B}^2$.

From Theorem \ref{indicator} we have $V_{C|B}\leq V^\mathrm{max}_{c_{C|B}}=\frac{4\pi}{3}(1-c_{C|B})^2$. Hence $\sqrt{V_{A|B}}+\sqrt{V_{C|B}}\leq\sqrt{\frac{4\pi}{3}}c_{C|B}+\sqrt{\frac{4\pi}{3}}(1-c_{C|B})=\sqrt\frac{4\pi}{3}$.}

\item The bound follows from the above result $\sqrt{V_{A|B}}+\sqrt{V_{C|B}}\leq\sqrt{\frac{4\pi}{3}}$ and the relationships $\gamma_b^4 V_{B|A} = \gamma_a^4 V_{A|B}$ and $\gamma_b^4 V_{B|C} = \gamma_c^4 V_{C|B}$, which apply for any state~\cite{Jevtic2013}. The bound is saturated for maximal volume $\EBA$ and $\EBC$ owing to Lemma \ref{mutual}, since the bound for Scenario (a) is saturated by maximal volume $\EAB$ and $\ECB$.
\end{myenumerate}
\end{proof}
\end{theorem}

These monogamy relations are remarkably elegant; it was not at all obvious \textit{a priori} that there would be such simple bounds for ellipsoid volume. The simplicity of the result is a consequence of the fact that $\EAB={\E}^{\max}_{c_{A|B}}$ implies $\EBA={\E}^{\max}_{c_{B|A}}$, $\ECB={\E}^{\max}_{c_{C|B}}$ and $\EBC={\E}^{\max}_{c_{B|C}}$, i.e. all of $\EAB$, $\EBA$, $\ECB$ and $\EBC$ are simultaneously maximal volume for their respective centres.

The monogamy of steering can easily be rephrased in terms of obesity. Although Alice and Bob's steering ellipsoid volumes are in general different, obesity is a party-independent measure. When expressed using obesity, the two steering scenarios therefore give the same bound $\Omega^2(\rho_{AB}) + \Omega^2(\rho_{BC}) \leq \gamma_b^{-2}$.

We now use the monogamy of steering to derive the Coffman-Kundu-Wootters (CKW) inequality for monogamy of concurrence~\cite{Coffman2000}. 

\begin{theorem}\label{ckw_inequality}
When Alice, Bob and Charlie share a pure three-qubit state the squared concurrences must obey the bound $C^2(\rho_{AB}) + C^2(\rho_{BC}) \leq 4 \det \rho_B$.
\begin{proof}
The result can be derived using either bound presented in Theorem \ref{volume_bounds}; we will use Scenario (a). Theorem \ref{concurrence_vol} tells us that $C(\rho_{AB})\leq \gamma_b^{-1}\big(\frac{3V_{A|B}}{4\pi}\big)^{1/4}$ and $C(\rho_{BC})\leq \gamma_b^{-1}\big(\frac{3V_{C|B}}{4\pi}\big)^{1/4}$, so that $\sqrt{\frac{4\pi}{3}}\gamma_b^2 C^2(\rho_{AB})\leq\sqrt{V_{A|B}}$ and $\sqrt{\frac{4\pi}{3}}\gamma_b^2 C^2(\rho_{BC})\leq\sqrt{V_{C|B}}$. The result then immediately follows from the bound  $\sqrt{V_{A|B}}+\sqrt{V_{C|B}}\leq\sqrt{\frac{4\pi}{3}}$ since $\gamma_b^{-2}=4\det \rho_B$.
\end{proof}
\end{theorem}

The monogamy of steering is strictly stronger than the monogamy of concurrence since Theorem \ref{volume_bounds} implies Theorem \ref{ckw_inequality} but not vice versa. Our derivation of the CKW inequality again demonstrates the significance of maximising steering ellipsoid volume for a given ellipsoid centre.

Finally, we note that the tangle of a three-qubit state may be written in the form $\tau_{ABC}=\gamma_b^{-2}-C^2(\rho_{AB})-C^2(\rho_{BC})$~\cite{Coffman2000}. When there is maximal steering, so that the bounds in Theorems \ref{volume_bounds} and \ref{ckw_inequality} are saturated, we have $\tau_{ABC}=0$. The corresponding three-qubit state belongs to the class of W states~\cite{Dur2000} (assuming that we have genuine tripartite entanglement). The W state itself, $\ket{W}=\frac{1}{\sqrt 3}(\ket{001}+\ket{010}+\ket{001})$, corresponds to the case that $\bi c_{A|B}=\bi c_{B|A}=\bi c_{C|B}=\bi c_{B|C}=(0,0,\frac{1}{2})$.

\section{Conclusions}

Any two-qubit state $\rho$ can be represented by a steering ellipsoid $\E$ and the Bloch vectors $\bi a$ and $\bi b$. We have found necessary and sufficient conditions for the geometric data to describe a physical two-qubit state $\rho \geq 0$. Together with an understanding of steering ellipsoid chirality, this is used to find the separable-entangled and physical-unphysical boundaries as a function of ellipsoid centre $\bi c$. These boundaries have geometric and physical significance. Geometrically, they can be used to find very non-trivial generalisations of Euler's inequality in classical Euclidean geometry. Physically, the maximal volume ellipsoids describe a family of states that are Choi-isomorphic to the amplitude-damping channel.

The concurrence of $\rho$ is bounded as a function of ellipsoid volume; this is used to show that maximal volume states also maximise concurrence for a given $\bi c$. By studying a system of three qubits we find relations describing the monogamy of steering. These bounds are strictly stronger than the monogamy of concurrence and provide a novel derivation of the CKW inequality. Thus the abstract, mathematical question of physicality and extremal ellipsoids naturally leads to an operational meaning for ellipsoid volume as a bound for concurrence and provides a new geometric perspective on entanglement monogamy.

These results may find applications in other notions of how `steerable' a state is. In particular, it should be possible to use our work to answer questions about EPR-steerable states~\cite{Wiseman2007}. For example, what are the necessary constraints on ellipsoid volume such that no local hidden state model can reproduce the steering statistics? Beyond this, the results on monogamy of steering pave the way for looking at steering in a many-qubit system by considering how bounds on many-body entanglement are encoded in the geometric data.

\ack

We wish to acknowledge useful discussions with Matthew Pusey. We are very grateful to Michael Hall for his assistance with the correction to the proof of Theorem 6(a). This work was supported by EPSRC and the ARC Centre of Excellence Grant No. CE110001027. DJ is funded by the Royal Society. TR would like to thank the Leverhulme Trust. SJ acknowledges EPSRC grant EP/K022512/1.

\appendix
\section*{Appendix}
\setcounter{section}{1}

Alice's steering ellipsoid $\E$ is invariant under Bob's local filtering transformation to the canonical frame as given in \eqref{eq:filtering}. We therefore need to consider only canonical states $\wt\rho$. Let us rephrase the conditions of Theorem \ref{conditions}, recalling that for $\E$ inside the Bloch sphere the condition $c^2 + \tr Q \leq 3$ is redundant: $\wt\rho\geq 0$ if and only if $g_1\geq 0\text{ and }g_2\geq 0$, where $g_1=c^4-2 u c^2 +q$ and $g_2=1 - \tr Q -2\chi\sqrt{\det Q}-c^2$. As discussed in Section \ref{section_canonical}, we can restrict our analysis to ellipsoids aligned with the coordinate axes, i.e. ellipsoids with a diagonal $Q$ matrix. Theorem~\ref{chirality} allows us to use the conditions for $\wt\rho\geq 0$ to probe both the separable-entangled and the physical-unphysical boundaries. Since all entangled $\E$ have $\chi=-1$, separable states lying on the separable-entangled boundary must correspond to the extremal ellipsoids that achieve $\wt\rho\geq 0$ with $\chi=+1$. Similarly, the physical-unphysical boundary corresponds to the extremal ellipsoids that achieve $\wt\rho\geq 0$ with $\chi=-1$. For the degenerate case with $\chi=0$, any physical ellipsoid must be separable and so the only boundary to find is physical-unphysical.

For a spherical $\E$ of radius $r$, centred at $\bi c$, we may set $Q=\mathrm{diag}(r^2, r^2, r^2)$ in the expressions for $g_1$ and $g_2$ to find the separable-entangled and physical-unphysical boundaries in $(c, r)$ parameter space. Similarly, for a circular $\E$ in the equatorial plane, we may set $Q=\mathrm{diag}(r^2, r^2, 0)$ and $c_3=0$ to find the physical-unphysical boundary.

More parameters are required to describe a general ellipse or ellipsoid, and so in these cases the procedure for finding extremal $\E$ is more involved. For an ellipsoid with semiaxes $\bi s$, the volume is $V=\frac{4\pi}{3}\sqrt{\det Q}=\frac{4\pi}{3}s_1 s_2 s_3$. We wish to maximise $V$ for a given $c$ subject to the inequality constraints $g_1\geq 0$ and $g_2 \geq 0$. This maximisation can be performed using a generalisation of the method of Lagrange multipliers known as the Karush-Kuhn-Tucker (KKT) conditions~\cite{Boyd2004}. We form the Lagrangian ${\cal L}=V+\lambda_1 g_1 + \lambda_2 g_2$, where $\lambda_1$ and $\lambda_2$ are KKT multipliers. Setting $Q=\mathrm{diag}(s_1^2, s_2^2, s_3^2)$, we then solve in terms of $c$ the system of equations and inequalities given by $\partial {\cal L}/\partial \bi {s}=\mathbf{0}$, $\lambda_1 g_1 = \lambda_2 g_2=0$ and $\lambda_1, \lambda_2, g_1, g_2\geq 0$. That the solution found corresponds to the global maximum is straightforwardly verified numerically.

This system can in fact be simplified before solving. In particular, the skew term $\bi{\hat c}^\mathrm{T}Q\bi{\hat c}$ is awkward to deal with in full generality. However, by symmetry, any maximal ellipsoid must have one of its axes aligned radially and the other two non-radial axes equal. Since we are looking at ellipsoids aligned with the coordinate axes, we may therefore take $\bi c=(0, 0, c)$ and $s_1=s_2$. Maximal solutions could then have $s_1=s_2>s_3$ (an oblate spheroid), $s_1=s_2<s_3$ (a prolate spheroid) or $s_1=s_2=s_3$ (a sphere).

Extremal ellipses in the equatorial plane are found similarly using the KKT conditions. We outline the method here in more detail as an example of the procedure. We describe an ellipse in the equatorial plane using $Q=\mathrm{diag}(s_1^2, s_2^2, 0)$ and $\bi c=(c, 0, 0)$. The area of this ellipse is $\pi s_1 s_2$. The Lagrangian is ${\cal L}=\pi s_1 s_2+\lambda_1 g_1 + \lambda_2 g_2$; the algebra is simplified by equivalently using ${\cal L}=8 \pi s_1 s_2+\lambda_1 g_1 + 2\lambda_2 g_2$. Substituting $Q$ and $\bi c$ into the expressions for $g_1$ and $g_2$ gives
\begin{eqnarray}
g_1 &= c^4-2c^2(1+s_1^2-s_2^2)+1-2s_1^2-2s_2^2-2s_1^2s_2^2+s_1^4+s_2^4,\label{eq:g1}\\
g_2 &= 1-s_1^2-s_2^2-c^2.
\end{eqnarray}

The requirement $\partial {\cal L}/\partial \bi {s}=\mathbf{0}$ corresponds to the two equations $\partial {\cal L}/\partial s_1=0$ and $\partial {\cal L}/\partial s_2=0$. Noting that the maximal solution must have $s_1\neq0$ and $s_2\neq0$, these equations can be solved simultaneously to give
\begin{eqnarray}
\lambda_1 &= \frac{s_2^2-s_1^2}{s_1 s_2(s_2^2-s_1^2+c^2)},\\
\lambda_2 &= \frac{1}{s_1 s_2}\left(s_1^2+s_2^2-\frac{s_2^2-s_1^2}{s_2^2-s_1^2+c^2}\right)\label{eq:l2}.
\end{eqnarray}

We now impose the constraints that $\lambda_1 g_1 = \lambda_2 g_2=0$ and $\lambda_1, \lambda_2, g_1, g_2\geq 0$. The only solution to this system of equations and inequalities requires $g_1=\lambda_2=0$. Using the expressions \eqref{eq:g1} and \eqref{eq:l2} these are solved simultaneously to find $s_1$ and $s_2$ in terms of $c$. Ruling out solutions that do not satisfy $0<s_1,s_2\leq 1$ gives the unique solution
\begin{eqnarray}
s_1=\frac{1}{4}(3-\sqrt{1+8c^2}),\\
s_2=\frac{1}{\sqrt{8}}\sqrt{1-4c^2+\sqrt{1+8c^2}},
\end{eqnarray}
as given in the main text.

\section*{References}
\bibliographystyle{iopart-num}
\bibliography{references}

\end{document}